# Field-induced spin polarization in lightly Cr-substituted layered antiferromagnet NiPS$_3$

Rabindra Basnet[1,2#], Dinesh Upreti[2], Taksh Patel[5], Santosh Karki Chhetri[2], Gokul Acharya[2], Md Rafique Un Nabi[2,3], Manish Mani Sharma[2], Josh Sakon[6], Mansour Mortazavi[1], Jin Hu[2,3,4*]

[1]Department of Chemistry & Physics, University of Arkansas at Pine Bluff, Pine Bluff, Arkansas 71603, USA

[2]Department of Physics, University of Arkansas, Fayetteville, Arkansas 72701, USA

[3]MonArk NSF Quantum Foundry, University of Arkansas, Fayetteville, Arkansas 72701, USA

[4]Materials Science and Engineering Program, Institute for Nanoscience and Engineering, University of Arkansas, Fayetteville, Arkansas 72701, USA

[5]Fayetteville High School, Fayetteville, Arkansas 72701, USA

[6]Department of Chemistry & Biochemistry, University of Arkansas, Fayetteville, Arkansas 72701, USA

Abstract

Tuning magnetic properties in layered magnets is an important route to realize novel phenomenon related to two-dimensional (2D) magnetism. Recently, tuning antiferromagnetic (AFM) properties through substitution and intercalation techniques have been widely studied in $MPX_3$ compounds. Interesting phenomena, such as diverse AFM structures and even the signatures of ferrimagnetism, have been reported. However, long-range ferromagnetic (FM) ordering has remained elusive. In this work, we explored the magnetic properties of the previously unreported Cr-substituted $NiPS_3$. We found that Cr substitution is extremely efficient in controlling spin orientation in $NiPS_3$. Our study reveals a field-induced spin polarization in lightly (9%) Cr-substituted $NiPS_3$, which is likely attributed to the attenuation of AFM interactions and magnetic anisotropy due to Cr doping. Our work provides a possible strategy to achieve FM phase in AFM $MPX_3$, which could be useful for investigating 2D magnetism as well as potential device applications.

[#] basnetr@uapb.edu

[*] jinhu@uark.edu

# I. Introduction

Tuning magnetic properties in two-dimensional (2D) magnetic materials provides opportunity for deeper understanding of low dimensional magnetism as well as enhance their feasibility for potential applications in next-generation devices [1–24]. Magnetic properties in layered magnetic materials have been found to be efficiently tunable by doping [25–56]. Particularly, introducing guest atoms in van der Waals (vdW) magnets can control the spin orientation, which leads to novel magnetic phenomenon arising from 2D magnetism. Metal thiophosphates $MPX_3$ ($M$ = metal such as Mn, Fe, Ni, etc; $X$ = chalcogen S and Se) compounds are such vdW magnets. Extensive efforts such as substitutions of metal ($M$) [25–40,56,57] or chalcogen ($X$) [46–50,57], and inter-layer intercalation [53–55,58] have been adopted to tune the antiferromagnetic (AFM) ground state in $MPX_3$. Various AFM structures have been obtained through $M$ and $X$ substitutions [25–40,46–50]. So far, whether ferromagnetism can be induced through substitution is still under debate. Theoretical efforts both supporting [59] and disapproving [60] this route have been reported. Experimentally, so far neither $M$ nor $X$ substitutions have been successful in stabilizing long-range ferromagnetic (FM) ordering in $MPX_3$. In addition to substitution, inter-layer intercalation is another doping strategy which is generally expected to cause charge doping. It has been proposed that ferromagnetism in $MPX_3$ can be induced by charge doping [61,62]. However, intercalating various guest species in $MPX_3$ have been reported to result in ferrimagnetism [53–55], yet ferromagnetism remains elusive.

Stabilizing ferromagnetism in $MPX_3$ is highly desired. First, it makes the study of 2D magnetism more accessible. That is because of the absence of net magnetization, direct probe of AFM ordering in atomically thin limit is challenging by using nanoscale magnetic characterization techniques such as scanning single-spin magnetometry [63], magneto-optical Kerr effect microscopy [1,2], polar reflective magnetic circular dichroism (RMCD) [64] and x-ray magnetic dichroism (XMCD) [65]. In addition, recent studies have predicted that FM phase of $MPX_3$ are not only limited within the boundary of 2D magnetism but could be extended to topological physics [59,66]. Hence, the rise of ferromagnetism in $MPX_3$ would provide a rare platform for investigating the interplay between magnetism and band topology in 2D. Furthermore, breaking time reversal symmetry by ferromagnetism may induce other exotic phenomena, especially in heterostacking of FM layered material and other quantum materials such as superconductors and topological materials, which require 2D FM materials.

This work focuses on tuning AFM properties by substituting new metal ions. Metal ion substitution in $MPX_3$ has led to a series of polymetallic "mixed" $(M_1)_{1-x}(M_2)_x PX_3$ compounds where $M_1$ and $M_2$ represent the original and substituted metal elements, respectively [25–40,56,67–70]. The previous substitution studies have mainly focused on three representative compounds, $MnPX_3$, $FePX_3$, and $NiPX_3$. Various metal ions such as $Mn^{2+}$ [28–30,33,36,38], $Fe^{2+}$ [28–30,32,33,37,40], $Co^{2+}$ [39], $Ni^{2+}$ [32,36–38,40], $Cu^{2+}$ [71], $Zn^{2+}$ [25–27,34,56,67], $Mg^{2+}$ [69], and $Cd^{2+}$ [68,70] have been adopted for substitution. Interestingly, Cr substitution in $MPX_3$ is still lacking despite recent developments in various Cr-based vdW magnets [1,2,4,8–10,12,14–18,51,52]. In this work, we have investigated Cr-substituted $NiPS_3$, and revealed field-induced spin polarization in 9% Cr-substituted $NiPS_3$. Such sensitive tuning in Cr-substituted $NiPS_3$ is likely attributed to weakening of AFM exchange and magnetic anisotropy due to Cr substitution that enables magnetic field to polarize Cr moments together with the partial polarization of some surrounding Ni ions. Our study offers a possible route to establish ferromagnetism in $MPX_3$, providing a novel platform to study 2D magnetism as well as an opportunity to develop magnetic materials-based next-generation devices.

## II. Experiment

The various $Ni_{1-x}Cr_xPS_3$ ($0 \leq x \leq 0.09$) single crystals used in this work were synthesized by a chemical vapor transport method using $I_2$ as the transport agent. Elemental powders with desired ratios were sealed in a quartz tube and heated in a two-zone furnace with a temperature gradient from 750 to 550 °C for a week. The elemental compositions and crystal structures of the obtained crystals were examined by energy-dispersive x-ray spectroscopy (EDS) and x-ray diffraction (XRD), respectively. Magnetization measurements were performed in a physical property measurement system (PPMS, Quantum Design).

## III. Result and Discussion

In $M$P$X_3$, the AFM ordering has been found to be extremely robust against perturbation from external magnetic field. The field-dependent studies of three representative compounds, $MnPS_3$, $FePS_3$, and $NiPS_3$, have revealed that a high magnetic field is needed to reorient moments of the two AFM sublattices into a spin-flop (SF) phase, which is manifested as the metamagnetic transition in the isothermal magnetization measurements under a strong field of $\mu_0H \approx 4$, 6, and 35 T applied along the magnetic easy axes at $T \approx 5$ K in $MnPS_3$ [72,73], $NiPS_3$ [38], and $FePS_3$ [74], respectively. After the SF transition, further applying the field leads to a linear increase of magnetization with no sign of saturation up to 20 and 9 T at $T \approx 5$ K in $MnPS_3$ [72] and $NiPS_3$ [38], respectively. In fact, a very high field up to ~ 40 T at a similar temperature is needed to saturate magnetization in $FePS_3$ [74]. This is suggestive of strong AFM interactions in $M$P$X_3$, which requires a high magnetic field to achieve a fully polarized FM phase.

In contrast, the magnetic moments in some layered AFM compounds can be easily polarized by applying a magnetic field. For example, Cr-based magnets such as $CuCrP_2S_6$ [75,76], $CrPS_4$ [77], $CrCl_3$ [78], and CrSBr [79] all exhibit AFM to field-polarized FM transitions at relatively low magnetic field below 8 T. Despite distinct crystal lattices, the Cr moments in these layered compounds are ordered in an A-type AFM configuration, which is characterized by antiferromagnetically coupled FM layers [77,80–83]. The interlayer AFM exchange between Cr moments is accompanied by essentially isotropic magnetic ordering because of small single-ion anisotropy for $Cr^{3+}$ ($d^3$) ion arising due to a quenched orbital moment [80,81]. Such weak AFM coupling and/or small magnetic anisotropy have been attributed to a field-driven AFM to FM transition at relatively low field in Cr-based AFM compounds [76,77,80,82]. This raises a natural question if Cr-based $MPX_3$ compounds could feature a similar low-field FM state.

The studies on $MPX_3$ compounds consisting of early 3$d$ transition metals such as Ti, V, and Cr have been very limited. So far, only $V_xPS_3$ ($x$ = 0.78 [84,85], 0.9 [86], and 1 [87]) and $CrPSe_3$ [88] have been reported. Indeed, our efforts in synthesizing the previously unreported $CrPS_3$ only yield another known compound $CrPS_4$. The difficulty in growing $CrPS_3$ has been proposed to originate from the destabilized P-P dimerization in $P_2S_6$ bipyramid structure unit, which is caused by a weaker Cr-S covalency in comparison to the metal-sulfur covalency in late-transition-metal-based $MPX_3$ ($M$ = Fe to Zn in the periodic table) compounds [89]. With the suppression of P-P dimerization, the $P_2S_6$ bipyramid unit changes valence from 4- to a nominal 6- in the ionic bonding approach, so that the divalent character for metal ion enhance to a higher valence state such as $Cr^{3+}$ and eventually favors the formation of $CrPS_4$ rather than $CrPS_3$ [89]. This scenario is in line with the fact that Cr ion usually form +3 valence in Cr-based vdW magnetic materials [75–77,79,90]. Similar scenario of +3 metal valence is also seen in another early-transition-metal-based $MPX_3$ compound $V_xPS_3$ ($x = 0.78$ [84,85] and 0.9 [86]), where $V^{2+}$ and $V^{3+}$ ions coexist and distribute randomly, giving rise to vacancies at the V sites [84–86].

Given the difficulty in direct synthesis for pristine $CrPS_3$, we switched to a different approach i.e., partial substitution of Cr in $NiPS_3$. In our study, we found that this is also challenging in comparison to other divalent metal ion substitutions [36–38,40,69,70]. Despite that various Cr contents in the source materials for synthesis have been tried, we were only able to introduce up to the highest 9% Cr in $NiPS_3$ [Fig. 1(a)]. This is in stark contrast to a much higher substitutions of Co [39], Cu [71], Mg [69], and Cd [68,70], and even a complete substitutions of Mn [30,36,38] and Fe [37,40] in $NiPS_3$. Such difficulty in doping Cr in $NiPS_3$ can be attributed to a similar reason that hinders the formation of pristine $CrPS_3$ phase as stated above.

Successful Cr doping has been demonstrated by composition analysis using EDS [Fig. 1(a)]. Given such a low amount of Cr, the compositions were determined after confirming homogeneous elemental phase in multiple locations of the crystals. We found that the highest Cr content can be obtained by our chemical vapor transport technique is 9% ($x = 0.09$), which limited our study to lightly doping regime. Nevertheless, as will be shown below, even light Cr-doping can induce strong modifications in magnetic properties in $NiPS_3$. To better understand the effect of Cr doping on magnetic properties, clarifying the doping scenario, i.e., substituting Ni or intercalating to the vdw gap, is important. The composition analysis by EDS quantitative analysis implies a substitution scenario without any metal vacancies, which is further supported by our structural analysis. The power XRD pattern obtained by grounding the single crystals is shown in Fig. 1(b). Cr substitution for Ni up to 9% (the highest Cr content obtained in this work) does not induce any structure transition. Only a tiny low-angle peak shift with Cr substitution is observed, as shown in the right panel of Fig. 1(b), which indicates a small expansion of $NiPS_3$ lattice. More structural information has been obtained via Rietveld refinement. We found a structural model [presented in Fig. 1(c)] assuming a substitution scenario and without vacancy sites nicely reproduce the powder XRD patterns. The extracted lattice constants are shown in Fig. 1(d), which reveals slight and systematic increases in all three lattice parameters $a$, $b$, and $c$ upon Cr substitution. Such increase is consistent with the tiny lattice expansion due to partial substitution of $Ni^{2+}$ by larger ion $Cr^{2+}$. On the other hand, an intercalation scenario can also expand the lattice. Generally, intercalating to the vdW gap generally causes a much stronger increase in $c$-axis as compared to $a$ and $b$, which has been observed in an earlier intercalation study on organic-ion intercalated $NiPS_3$ [58]. However, such anisotropic lattice expansion is very different from the nearly isotropic lattice expansion along all three axes in our samples, as shown in Fig. 1(d).

In addition to identifying a substitution scenario for Cr doping, the above composition and structure analyses also demonstrate the absence of metal vacancies, which supports a +2 valence for the substituted Cr ions. Clarifying the Cr valence is important because it determines the $d$-orbital occupancies that affect the magnetic exchange interaction. In addition, the presence of vacancies can cause local structure distortions that affect local magnetism, creating additional complexity in understanding the mechanisms of the modification of magnetism by doping.

In $NiPS_3$, though the light metal substitution does not significantly modify the lattice type, it still influences the magnetic properties. For example, substituting 5% Mn [38] and 10% Fe [37] can efficiently control the magnetism in $NiPS_3$. To investigate the effects of Cr substitution on magnetic properties of $NiPS_3$, we measured the field $M(H)$ and temperature $M(T)$ dependences of magnetization for our lightly substituted $Ni_{1-x}Cr_xPS_3$ samples. As shown in $M(H)$ measurements under in-plane [$H//ab$, Fig. 2(a)] and out-of-plane [$H \perp ab$, Fig. 2(b)] magnetic fields at $T = 2$ K, substituting up to 9% Cr for Ni has substantial effect on field dependence of magnetization, resulting in systematic enhancement of magnetization with Cr substitution. In addition, the SF transition is strongly modified in Cr-substituted NiPS$_3$, which will be discussed below. The tunable magnetism in NiPS$_3$ is also demonstrated by the composition dependence of Néel temperature $T_N$. Fig. 3(a) presents the temperature dependence of susceptibility under in-plane ($H//ab$, blue) and out-of-plane ($H\perp ab$, red) magnetic fields of $\mu_0 H = 0.1$ T. To obtain the precise $T_N$, we calculated the derivative $d\chi/dT$ for susceptibility data and used their peak position to define $T_N$ [Inset, Fig. 3(a)], which has been widely used in previous studies [36,37,40,49]. The extracted $T_N$ for the parent compound NiPS$_3$ is found to be 155 K, consistent with the reported values [38,91]. As summarized in Fig. 4 (Top panel), in Ni$_{1-x}$Cr$_x$PS$_3$, the $T_N$ significantly drops from 155 K for $x = 0$ to 34 K up on only 9% Cr-substitution.

Highly sensitive magnetism to Cr substitution in NiPS$_3$ is further supported by the evolution of SF transition with Cr substitution. NiPS$_3$ exhibits a metamagnetic transition at an in-plane magnetic field of 6 T [Fig. 3(b)], which is characterized by a clear upturn (shown by blue arrows) in isothermal magnetizations $M(H)$ at $T = 2$ K. Such metamagnetic transition under an in-plane field in NiPS$_3$ has recently been discovered and was attributed to a SF transition [38], consistent with the nearly in-plane magnetic moment orientation for Ni ions in this compound [91]. The SF behavior of NiPS$_3$ is significantly modified upon Cr substitution. As shown in Fig. 3(b), with increasing Cr content $x$, the SF transition, which occurs under an in-plane field ($H//ab$), is strongly suppressed, as manifested by the less-obvious magnetization upturn and reduced SF field ($H_{SF}$) from $\mu_0 H_{SF} \approx 6$ T (for $x = 0$) to 0.6 T (for $x = 0.09$) at $T = 2$ K. Furthermore, the SF transition is observed only under in-plane field (blue arrow) in $x = 0$-0.035 samples while it appears under both in-plane (blue arrow; $\mu_0 H_{SF} \approx 0.6$ T) and out-of-plane (red arrow; $\mu_0 H_{SF} \approx 0.8$ T) magnetizations in the $x = 0.09$ sample. The tuning of SF transition is summarized in Fig. 4 (Middle panel), which suggests the moment reorientation upon Cr substitution as will be discussed later.

In NiPS$_3$, the SF state is driven by external magnetic field along magnetic easy axis, which rotate the Ni moments and causes weak FM component arising from the uncompensated canted moments [38]. Applying field exceeding $H_{SF}$ and up to 9 T is unable to polarize such canted moments in NiPS$_3$ ($x = 0$) (Fig. 2), which is consistent with the previous report [38]. Also, though substituting Cr for Ni suppresses the $H_{SF}$ in the $x = 0.012$ sample (Fig. 4), its magnetization does not saturate up to 9 T. However, this scenario appears to change completely with further increasing Cr amount. For $x = 0.035$ and 0.09 samples, the SF transitions occur at much lower $H_{SF}$, and a moment polarization behavior featuring sublinear magnetization starts to occur at higher fields. In particular, a clear kink near 8 T can be observed in both in-plane [Fig. 2(a)] and out-of-plane [Fig. 2(b)] magnetizations of the $x = 0.09$ sample, as indicated by the black arrows in the figures, which indicates spin polarization. Above the saturation field of $H_{sat} \sim 8$ T at $T = 2$ K for $x = 0.09$, the magnetization for both field directions attains a value ~0.24 $\mu_B$ per f.u. [Figs. 2(a and b)]. Considering the moment of Ni$^{2+}$ of 2.82 $\mu_B$ and the moment of Cr$^{2+}$ of 4.60 $\mu_B$, substituting 9% Ni with Cr results in a theoretical magnetic moment ~ 2.98 $\mu_B$ per f.u, as denoted by the blue dashed lines in Fig. 2. The observed much lower saturation moment indicates that a full spin polarization is not achieved. Such spin polarization behavior can be understood in terms of the polarization of only Cr and the surrounding Ni ions. Substitution of a Cr$^{2+}$ for a Ni$^{2+}$ in an AFM lattice would result in net magnetic moment of $\mu$ (Cr$^{2+}$) - $\mu$ (Ni$^{2+}$) = 4.60 $\mu_B$ - 2.82 $\mu_B$ = 1.78 $\mu_B$. Therefore, the 9% Cr$^{2+}$ substitution should yield a net moment of ~0.16 $\mu_B$ per f.u. This, together with the partial polarization of some surrounding Ni ions, a total net moment of 0.24 $\mu_B$ per f.u. is reasonable.

To understand the effect of Cr substitution, we performed systematic isothermal magnetization measurements at various temperatures for the $x = 0.09$ sample. As shown in Figs. 5(a) and 5(b), the in-plane and out-of-plane isothermal magnetizations display a saturation-like behavior above $H_{sat} \sim 8$ T at $T = 2$ K. The saturation field $H_{sat}$ is gradually reduced with rising temperature. The reduction of $H_{sat}$ can be ascribed to the additional thermal energy upon heating that assist moment polarization under lower field. Though increasing temperature favors moment polarization, high temperature also randomizes magnetic moments and attenuates the correlation between them. Therefore, the SF transition is no longer clearly visible, and the field-dependent isothermal magnetization gradually evolves from a sublinear to a linear field dependence above 30 K. In fact, this temperature also corresponds to the AFM ordering temperature $T_N$, which is suppressed from 155 K in pristine NiPS$_3$ [38] to 34 K in this 9% Cr substituted sample, as mentioned earlier. Compared to other metal ion substitution in NiPS$_3$ such as Mn substitution for Ni [38], Cr substitution is much more efficient in suppressing magnetic ordering temperature. Figures 5(c) and 5(d) show temperature dependent in-plane and out-of-plane magnetic susceptibilities, respectively, for the identical $x = 0.09$ sample measured under various magnetic fields from 0.1 T to 9 T. At 0.1 T, $T_N \approx 34$ K characterized by a weak susceptibility peak, and more precisely in differential susceptibility $d\chi/dT$ [Inset of Fig. 3(a)] shifts to a lower temperature upon increasing magnetic field, and eventually becomes unobservable down to $T = 2$ K above 3 T. Such suppression of AFM ordering by relatively low field for Cr-substituted NiPS$_3$ is in stark contrast to the field-independent $T_N$ for pristine NiPS$_3$ [38], implying weakened AFM coupling due to Cr substitution. The weakening of AFM interaction is also in line with the field-driven spin polarization at relatively low field in $x = 0.09$ sample, as seen in other Cr-based antiferromagnets [76,77,80,82].

In addition to weak AFM interactions, a low magnetic anisotropy has also been proposed to be a possible reason for the field-induced moment polarization in some Cr-based AFM compounds [76,77,80,82]. The strength of magnetic anisotropy can be evaluated from the evolution of SF behavior or vice-versa. One way to clarify it is the comparison of SF transition in (Mn,Ni,Fe)PS$_3$, in which $H_{SF}$ increases with enhancing single-ion anisotropy from MnPS$_3$ to FePS$_3$ in the order MnPS$_3$ < NiPS$_3$ < FePS$_3$ (4, 6, and 35 T, respectively, at $T \sim$ 5K) [38,73,74]. Furthermore, the modification of magnetic anisotropy in $M$P$X_3$ compounds by metal substitution has been reported to tune SF transition [27,38]. For example, in Zn-substituted MnPS$_3$ [27], 20% Zn substitution for Mn reduces $H_{SF}$ by half, which has been ascribed to the weakening of magnetic anisotropy due to diluting Mn magnetic lattice by non-magnetic Zn substitution. Such tunable SF transition in MnPS$_3$ is expected given negligible single-ion anisotropy for Mn ($A \approx$ 0.0086 meV) [92]. For NiPS$_3$, owing to the more pronounced $A$ ($\approx$ 0.3 meV) for Ni [93], the tuning of SF transition could be more difficult than MnPS$_3$. However, surprisingly, here we found that the SF transition in NiPS$_3$ is sensitive to very light Cr substitution. In fact, similar highly sensitive SF transition to light substitution has also been discovered in our earlier study on Mn-substituted NiPS$_3$ [38]. As mentioned earlier, the SF field $H_{SF}$ monotonically decreases with increasing Cr content $x$ in Ni$_{1-x}$Cr$_x$PS$_3$ (Fig. 4). These observations suggest reduced magnetic anisotropy by Cr substitution in NiPS$_3$, which is supported by the evolution of the effective magnetic anisotropy field. The spin-flop field can be approximately expressed as $H_{SF} \approx \sqrt{2H_E H_A}$, where $H_E$ and $H_A$ are the effective fields characterizing magnetic exchange and anisotropy, respectively [27,94]. So, the anisotropy field $H_A$ can be estimated by $H_A \approx (H_{SF})^2/(H_E)$. The spin-flop field $H_{SF}$ can be evaluated from the field-dependent magnetization, and effective exchange field $H_E$ can be assessed by ordering temperature $T_N$. From the temperature-dependent susceptibility [Fig. 3(a)] and field-dependent magnetization [Fig. 3(b)] measurements, we have extracted the composition dependence of magnetic ordering temperature $T_N$ and spin-flop field $H_{SF}$ at 2K, as summarized in Fig. 4. With Cr substitution up to 9%, $T_N$ is suppressed by a factor of 4.6 from ~155 K to ~ 34 K, while $H_{SF}$ decreases by a factor of 10 from 6 T to 0.6 T. This leads to a significant suppression of the anisotropy field $H_A$: $H_A$ in the $x = 0.9$ sample is reduced to only 5% of that of the undoped sample. Such result indicates Cr substitution is very efficient in suppressing the magnetic anisotropy. In addition, reduced magnetic anisotropy is also supported by the evolution of low field (below $H_{SF}$) magnetization. As shown in Fig. 3(b), increasing Cr content $x$ gradually suppresses anisotropy between in-plane and out-of-plane magnetizations below $H_{SF}$. The reduced magnetic anisotropy with Cr substitution generates more controllable moment orientation that eventually polarizes towards field direction as seen in $x = 0.09$ in $Ni_{1-x}Cr_xPS_3$.

In $MPX_3$, tuning magnetic anisotropy can modify the moment orientations. As mentioned earlier, the weakening of magnetic anisotropy on increasing Cr substitution implies the rotation of moments from the nearly in-plane orientation in pristine $NiPS_3$ towards the out-of-plane direction, as illustrated in the schematic in Fig. 6. Increasing Cr content $x$ causes the spin to rotate away from the basal plane. Because the SF transition in AFM compounds is driven by magnetic field component along the magnetic easy axis, the absence of the SF transition under out-of-plane magnetic field in $x = 0.012$ and $0.036$ samples [Fig. 4(a)] implies that the canted moments should not deviate too much from the basal plane with less amount of Cr substitution, which is similar to pristine $NiPS_3$. Further increasing the Cr content to 9% (i.e., $x = 0.09$), weak but clear metamagnetic transitions appear with the application of both in-plane and out-of-plane magnetic fields [Figure 3(b)], which is in stark contrast to metamagnetic transition observed only along the magnetic easy axis in Cr-based antiferromagnets [75–79]. Such distinct SF behavior in the $x = 0.09$ sample might be attributed to strongly canted moments. Having said that, the slightly lower in-plane $H_{SF}$ in $x = 0.09$ sample [Fig. 4(b)] suggests that the moments could still be tilted slightly towards the *ab*-plane. With such canted AFM ordering, both in-plane and out-of-plane fields can reorient moments to a SF state above some $H_{SF}$, as shown in Fig. 6. Further increasing the field (i.e., $H_{SF} \leq H \leq H_{sat}$) will gradually polarize the canted moments above a saturation field $H_{sat}$.

The above results demonstrate that Cr substitution can strongly tune the magnetic properties in $NiPS_3$. We also want to emphasize that the lattice of $NiPS_3$ barely changes upon Cr substitution [Figs. 1(b and d)]. Such sensitivity to light Cr substitution (a few percent) in $NiPS_3$ is quite interesting and demands further theoretical and experimental investigations to clarify the complete mechanism. Furthermore, our study sheds light into a few important questions that have so far remained elusive. For example, *why do all the Cr-based antiferromagnets exhibit similar magnetic properties such as low-field spin-flop transition followed by spin polarization at high field? How to incorporate more Cr into $MPX_3$ lattice?* Furthermore, another previously largely overlooked question in many doped systems is that, *will those dopants form possible long-range or short-range ordering to further affect the properties?* Answering those questions could be important for future material development and implementations, and doped $MPX_3$ systems provide an excellent platform to investigate and address those important question.

In conclusion, we have studied the previously unexplored Cr-substituted $NiPS_3$ compounds. In contrast to other metal-ion substitutions such as Mn, Fe, Co, Cu, and Zn, Cr substitution in $NiPS_3$ is found to be very challenging and light substitution up to only a few percent can be realized. We found that Cr substitution can significantly alter the magnetic properties of $NiPS_3$. Such doping effects in $NiPS_3$ can be understood in terms of tuning magnetic exchange and anisotropy, and eventually results in a field-driven spin polarization in the $Ni_{0.91}Cr_{0.09}PS_3$ sample.

This finding suggests that Cr doping could be a possible route to achieve FM state in $M$P$X_3$. Furthermore, our work also provides a novel platform to study 2D magnetism in $M$P$X_3$ and design nanodevices based on magnetic materials for spintronics applications.

**Acknowledgement**

Sample synthesis and property measurements were supported by the U.S. Department of Energy, Office of Science, Basic Energy Sciences, under Award No. DE-SC0022006. XRD refinement and anisotropic field analysis were supported by μ-ATOMS, an Energy Frontier Research Center funded by DOE, Office of Science, Basic Energy Sciences, under Award DE-SC0023412. J. S. acknowledges the support from NIH under award P20GM103429 for powder XRD.

**Figure 1**

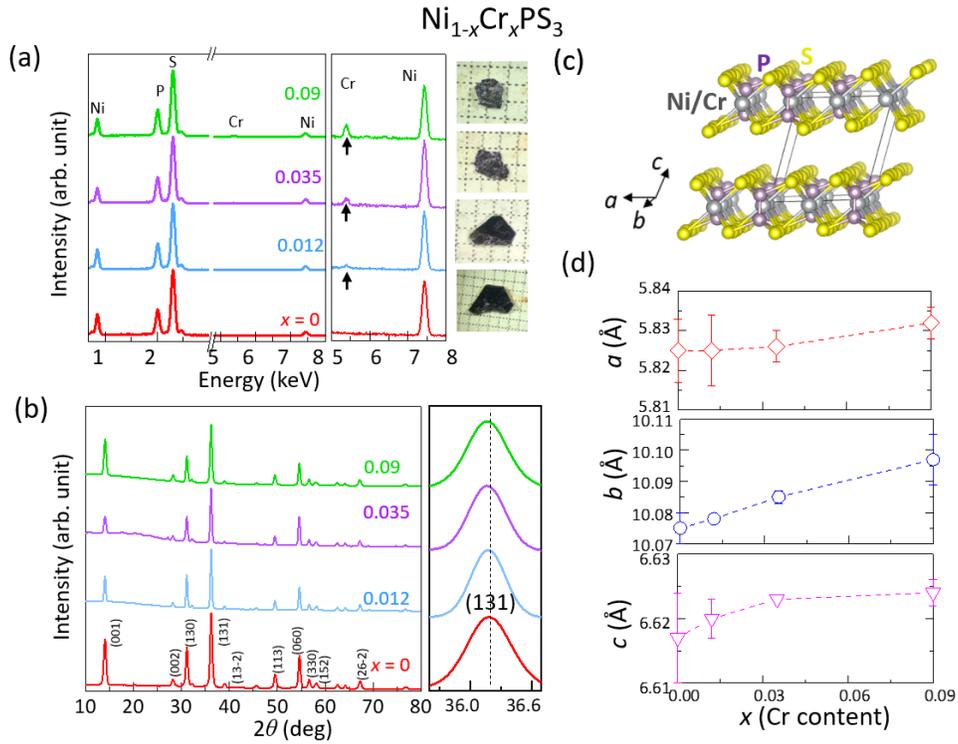

FIG. 1. (a) Energy dispersive x-ray spectroscopy (EDS) along with crystal images. The right panel shows the increased EDS spectrum intensity for Cr as denoted by the black arrows. (b) X-ray diffraction (XRD) result for Cr-substituted NiPS$_3$, Ni$_{1-x}$Cr$_x$PS$_3$ ($0 \leq x \leq 0.09$). Right panel in Figs. 1(b) shows evolution of XRD peak in which the dashed lines are guide to eye. (c) The crystal structure of Ni$_{1-x}$Cr$_x$PS$_3$ ($0 \leq x \leq 0.09$). (d) Composition dependence of lattice parameters *a*, *b*, and *c* in Ni$_{1-x}$Cr$_x$PS$_3$ ($0 \leq x \leq 0.09$).

**Figure 2**

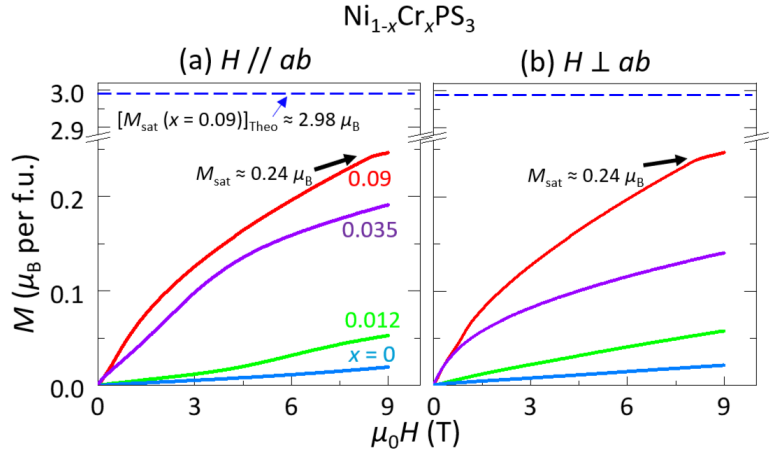

FIG. 2. Field dependence of magnetization for $Ni_{1-x}Cr_xPS_3$ ($0 \leq x \leq 0.09$) samples at $T = 2$ K under (a) in-plane ($H \parallel ab$) and (b) out-of-plane ($H \perp ab$) magnetic fields. The black arrows denote saturation of magnetization for the $x = 0.09$ sample. The dashed lines denote the theoretical saturated magnetization value of 2.98 $\mu_B$ for the $x = 0.09$ sample, if all metal ion moments are polarized.

**Figure 3**

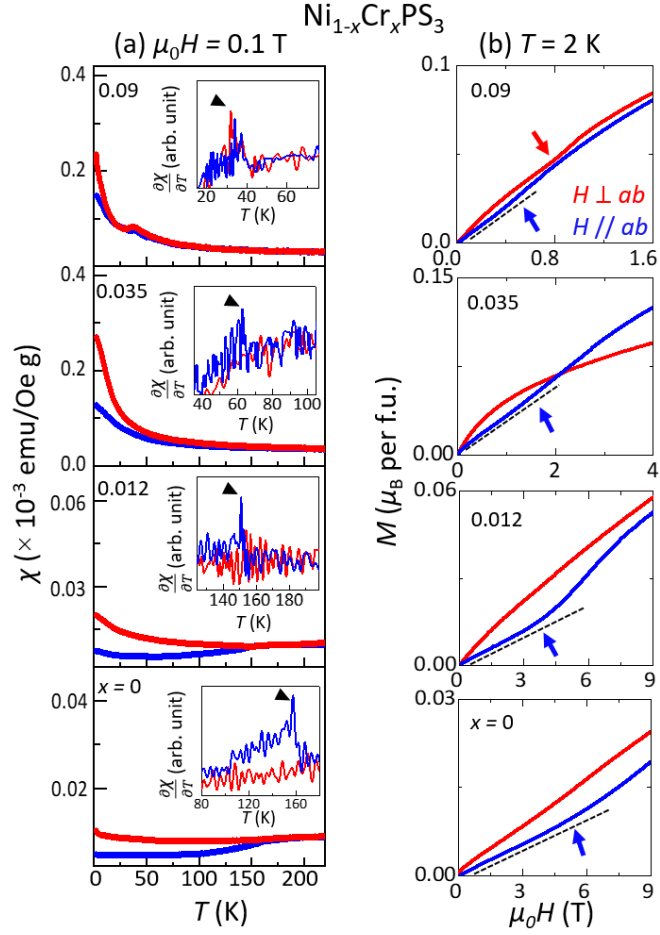

FIG. 3. (a) Temperature dependence of susceptibility for $Ni_{1-x}Cr_xPS_3$ ($0 \leq x \leq 0.09$) samples under in-plane ($H\|ab$, blue) and out-of-plane ($H\perp ab$, red) magnetic fields of $\mu_0 H = 0.1$ T. Insets: Temperature dependence of derivative susceptibility $d\chi/dT$. The peak of $d\chi/dT$ denotes $T_N$, which is denoted by black triangles. (b) Field dependence of magnetization for $Ni_{1-x}Cr_xPS_3$ ($0 \leq x \leq 0.09$) samples at $T = 2$ K under out-of-plane ($H\perp ab$, red) and in-plane ($H\|ab$, blue) magnetic fields. The red and blue arrows in fig. (b) denote spin-flop field under $H\perp ab$ and $H\|ab$ magnetic fields, respectively. The dashed lines are guide to eye.

**Figure 4**

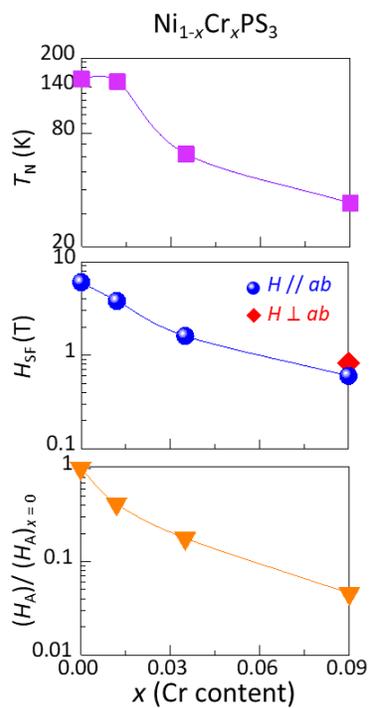

FIG. 4. Doping dependence of Néel temperature ($T_N$), spin-flop field ($H_{SF}$), and the ratio of effective magnetic anisotropy field ($H_A$) of Ni$_{1-x}$Cr$_x$PS$_3$, normalized to $H_A$ for the $x = 0$ sample.

**Figure 5**

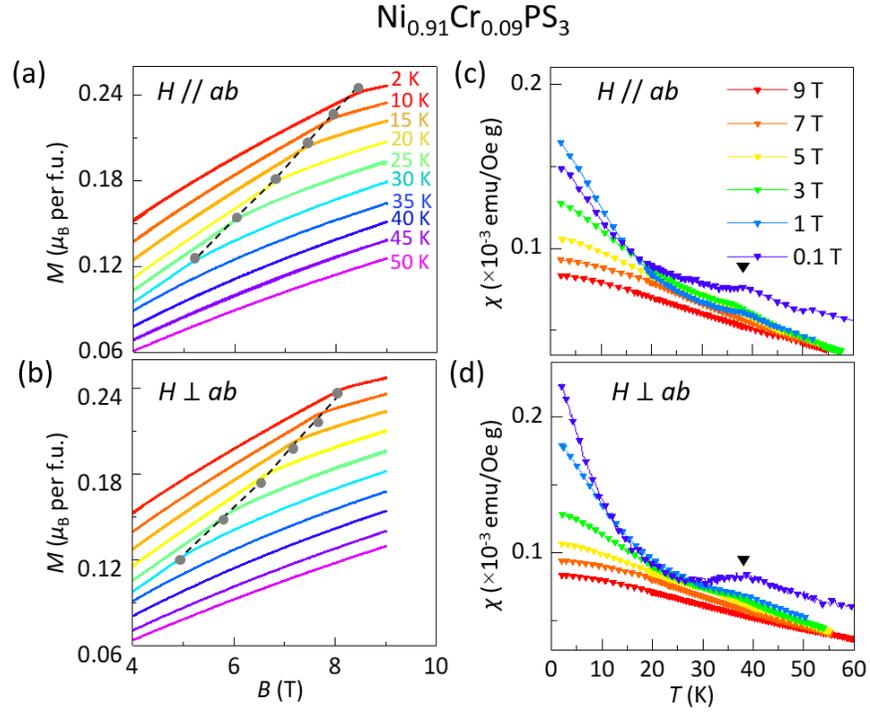

FIG. 5. (a-b) Field dependence of magnetization for $Ni_{0.91}Cr_{0.09}PS_3$ sample at various temperatures from 2 to 50 K under (a) in-plane ($H\|ab$) and (b) out-of-plane ($H\perp ab$) magnetic fields. The dashed lines denote the evolution of saturation field ($H_{SF}$) with temperature. (c-d) Temperature dependence of susceptibility for $Ni_{0.91}Cr_{0.09}PS_3$ sample at various (c) in-plane ($H\|ab$) and (d) out-of-plane ($H\perp ab$) magnetic fields from 0.1 to 9 T. The black triangles denote $T_N$.

**Figure 6**

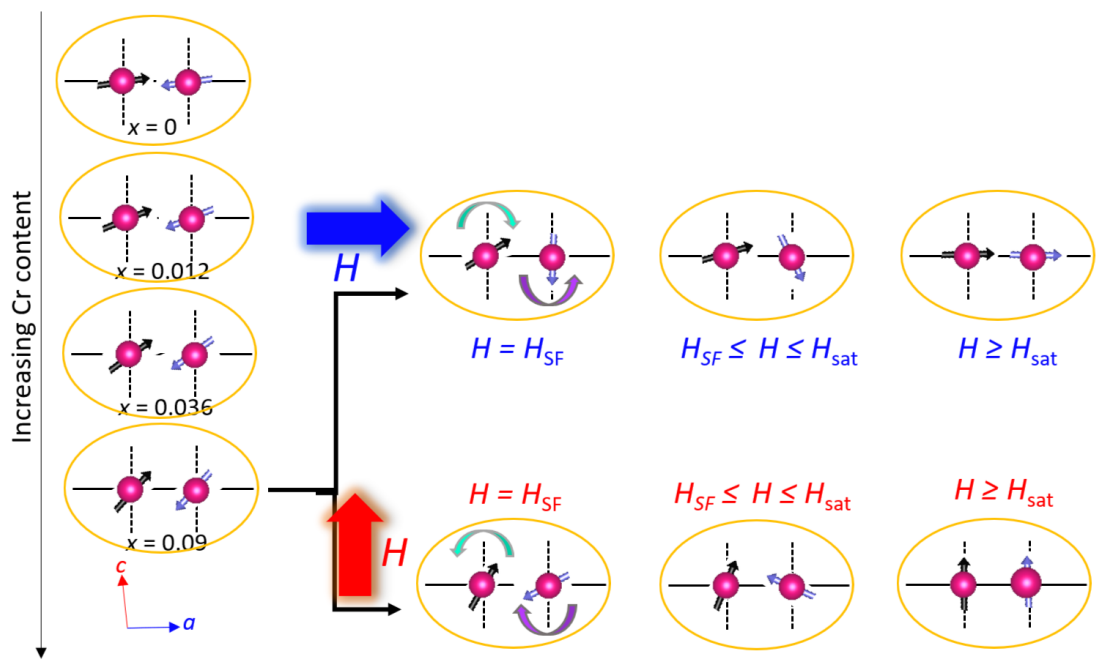

FIG. 6 Schematic of evolution of magnetic moment orientations with Cr substitution up to 9% in NiPS$_3$. This schematic further displays evolution of magnetic moments under in-plane ($H\|ab$) and out-of-plane ($H\perp ab$) magnetic fields in Ni$_{0.91}$Cr$_{0.09}$PS$_3$ sample.